\documentclass[twocolumn,showpacs,preprintnumbers,amsmath,amssymb]{revtex4}

\usepackage{graphicx}
\usepackage{dcolumn}
\usepackage{bm}

\newcommand{\bq}{\begin{equation}}
\newcommand{\eq}{\end{equation}}
\newcommand{\bqa}{\begin{eqnarray}}
\newcommand{\eqa}{\end{eqnarray}}
\newcommand{\nn}{\nonumber \\}

\def\be     {\begin{equation}}
\def\ee     {\end{equation}}
\def\bea        {\begin{eqnarray}}
\def\eea        {\end{eqnarray}}
\def\bnn    {\begin{eqnarray*}}
\def\enn    {\end{eqnarray*}}

\begin{document}

\title{Critical field theory of the Kondo lattice model in two dimensions}
\author{Ki-Seok Kim }
\affiliation{Korea Institute of Advanced Study, Seoul 130-012,
Korea}
\date{\today}

\begin{abstract}
In the context of the U(1) slave boson theory we derive a critical
field theory near the quantum critical point of the Kondo lattice
model in two spatial dimensions. First we argue that strong gauge
fluctuations in the U(1) slave boson theory give rise to
confinement between spinons and holons, thus causing "neutralized"
spinons in association with the slave boson U(1) gauge field.
Second we show that critical fluctuations of Kondo singlets near
the quantum critical point result in a new U(1) gauge field. This
emergent gauge field has nothing to do with the slave boson U(1)
gauge field. Third we find that the slave boson U(1) gauge field
can be exactly integrated out in the low energy limit. As a result
we find a critical field theory in terms of renormalized
conduction electrons and neutralized spinons interacting via the
new emergent U(1) gauge field. Based on this critical field theory
we obtain the temperature dependence of specific heat and the
imaginary part of the self-energy of the renormalized electrons.
These quantities display non-Fermi liquid behavior near the
quantum critical point.
\end{abstract}


\maketitle

It is now believed that the classical critical field theory of
order parameter fluctuations, the Hertz-Millis theory of
paramagnons in the Landau-Ginzburg-Wilson ($LGW$) theoretical
framework, cannot explain the observed non-Fermi liquid behavior
in thermal and electrical properties in the heavy fermion
metals\cite{Si}. Although the $LGW$ theoretical framework is the
cornerstone of the theory of phase transitions, it is now
necessary to go beyond the $LGW$ theory. Recently Si et al.
claimed that the non-Fermi liquid behavior can be explained by
local quantum criticality\cite{Si}. According to their claim,
electronic excitations resulting from the Kondo resonances
participate in the critical theory, while in the traditional
critical theory the fermionic excitations are just
bystanders\cite{Si}. Another approach is based on an exotic phase
of the U(1) slave boson theory, so called a deconfinement phase in
the context of the gauge theory\cite{Senthil}. In the U(1) slave
boson theory the critical theory is written in terms of not only
order parameter fluctuations (holons) but also fermionic
excitations (spinons). Furthermore, these spinons and holons
interact via long range interactions mediated by the internal
(slave boson) U(1) gauge fields. This critical theory successfully
explains the non-Fermi liquid behavior in the Kondo
problem\cite{Senthil}.

In the present paper we reexamine the critical field theory in the
U(1) slave boson representation. Despite the successful
description of the non-Fermi liquid behavior\cite{Senthil} it is
still controversial, especially in two dimensions. This is due to
the fact that the U(1) gauge field is compact, thus admitting
instanton excitations. In two dimensions the instanton excitations
of the U(1) gauge field are believed to result in
confinement\cite{Fradkin} of the slave particles except some
special
cases\cite{Senthil,Fradkin,Deconfinement1,Deconfinement2,Deconfinement3,Kim_PRL}.
The slave boson critical theory does not belong to any special
cases of the previous
studies\cite{Senthil,Fradkin,Deconfinement1,Deconfinement2,Deconfinement3,Kim_PRL}.
If the confinement occurs, it is necessary to find a new critical
field theory in terms of internal charge "neutral" particles
emerging from the confinement. In the context of high $T_c$
cuprates it has been also pointed out that in the slave boson mean
field theories strong gauge fluctuations mediating interactions
between the slave particles are not appropriately taken into
account\cite{Nayak,DHLee}.

In this paper we derive a new critical field theory near the
quantum critical point of the Kondo lattice model in two spatial
dimensions. We find that the critical field theory can be written
in terms of internal charge neutral fermions interacting via new
emergent U(1) gauge fields [Eq. (11)]. The neutral fermions result
from the confinement via the internal U(1) gauge field. The new
emergent U(1) gauge field arises from critical fluctuations of
Kondo singlets near the quantum critical point. Investigating the
temperature dependence of specific heat and the imaginary part of
the self-energy of the neutral fermions, we show that this new
critical field theory can explain the anomalous behavior near the
quantum critical point.

We consider the two dimensional Kondo lattice model in the U(1)
slave boson representation \bqa && Z =
\int{Dc_{\sigma}}{Df_{\sigma}}{Da_{\tau}}
e^{-\int_{0}^{\beta}{d\tau} L} , \nn && L =
\sum_{k}c_{k\sigma}^{\dagger}(\partial_{\tau} -
\epsilon_{k})c_{k\sigma} + \frac{J_K}{2}\sum_{r}{\bf
S}_{r}\cdot{c}_{r\sigma}^{\dagger}\tau_{\sigma\sigma'}c_{r\sigma'}
\nn && + \sum_{r}f_{r\sigma}^{\dagger}(\partial_{\tau} -
ia_{\tau})f_{r\sigma} + J_{H}\sum_{<r,r'>}{\bf S}_{r}\cdot{\bf
S}_{r'} . \eqa Here $c_{\sigma}$ is the conduction electron and
$f_{\sigma}$, the spinon describing spin degrees of freedom of
localized electrons. $\sigma$ ranges over spin $\uparrow$ and
$\downarrow$. ${\bf S}_{r} =
\frac{1}{2}f_{r\sigma}^{\dagger}\tau_{\sigma\sigma'}f_{r\sigma'}$
is the localized spin. $\epsilon_{k}$ is the bare dispersion of
the conduction electrons. $a_{\tau}$ is a Lagrange multiplier to
guarantee a single occupancy constraint for spinons. It can be
considered to be the time component of a gauge field as we will
see later. $J_{K}$ is the Kondo coupling constant between the
conduction and localized spins, and $J_H$, the antiferromagnetic
Heisenberg coupling constant between localized spins.

Following Senthil et al.\cite{Senthil}, we obtain a one-body
effective Lagrangian in terms of the conduction electrons and
spinons coupled to order parameters \bqa && Z =
\int{Dc_{\sigma}}{Df_{\sigma}}{Db}{Da_{\mu}}e^{-\int_{0}^{\beta}{d\tau}
L} , \nn && L = \sum_{k}c_{k\sigma}^{\dagger}(\partial_{\tau} -
\epsilon_{k})c_{k\sigma} \nn && +
\sum_{r}f_{r\sigma}^{\dagger}(\partial_{\tau} -
ia_{\tau})f_{r\sigma} -
\chi_{0}\sum_{<r,r'>}(f^{\dagger}_{r\sigma}e^{ia_{rr'}}f_{r'\sigma}
+ h.c.) \nn && -
\sum_{r}(b^{\dagger}_{r}f_{r\sigma}c_{r\sigma}^{\dagger} + h.c.) +
\sum_{r}\frac{4|b_r|^2}{J_K} .  \eqa Here $b_{r}$ is the holon
which represents hybridization between the conduction electron and
spinon. $\chi_{rr'} = \chi_{0}e^{ia_{rr'}}$ is the hopping order
parameter of the spinons where $a_{rr'}$ is the internal U(1)
gauge field originating from a composite field (slave boson)
representation. The U(1) gauge field mediates long range
interactions between the spinons and holons\cite{Senthil}. Mean
field values of the order parameters are given by\cite{Senthil}
\bqa && b^{\dagger}_{r} =
\frac{J_{K}}{2}<c_{r\sigma}^{\dagger}f_{r\sigma}> , \nn &&
\chi_{0} = \frac{J_{H}}{2}<f_{r\sigma}^{\dagger}f_{r'\sigma}> .
\eqa Here we consider the case of half filling for the localized
electrons as mentioned above, i.e., $1 =
<f_{r\sigma}^{\dagger}f_{r\sigma}>$.

In passing, let us discuss mean field phases. In strong Kondo
coupling regime the conduction electrons and spinons are strongly
hybridized and thus the Kondo singlets are expected. This is
represented by holon condensation, i.e.,
$<c_{r\sigma}^{\dagger}f_{r\sigma}> \not= 0$. Magnetic moments of
the localized electrons are screened by conduction electrons and
the local moments are expected to disappear. In this case not only
the conduction electrons but also the spinons participate in
conduction via the hybridization. Thus Fermi liquid with large
Fermi surface is expected to occur\cite{Senthil}. In weak Kondo
coupling regime the hybridization vanishes and the Kondo singlets
disappear. The holons are not condensed, i.e.,
$<c_{r\sigma}^{\dagger}f_{r\sigma}> = 0$. The local magnetic
moments are not screened and they are expected to appear via the
antiferromagnetic coupling $J_H$. In this case the conduction
electrons are decoupled from the spinons and only the conduction
electrons participate in conduction\cite{Senthil}. Thus Fermi
liquid with small Fermi surface is expected to
appear\cite{Senthil}.

At the critical Kondo coupling the hybridization is strongly
fluctuating, causing critical fluctuations of the Kondo singlets.
One may obtain a critical field theory for the order parameter
(holon) fluctuations. New constituents in the critical field
theory are fermionic excitations, here the spinons. The spinons
interact with the holons via long range gauge interactions.
Integrating over the conduction electrons, one can obtain the
critical theory in terms of the spinons and holons interacting via
the U(1) gauge fields near the quantum critical
point\cite{Senthil}. It is important to notice that the U(1) gauge
field is compact. If the compactness is ignored, this critical
theory successfully explains the non-Fermi liquid behavior in the
heavy fermion metals\cite{Senthil}. However, as discussed earlier,
it is generally accepted that owing to the instanton excitations
originating from the compact U(1) gauge field only the confinement
phase is expected to exist in two dimensions\cite{Fradkin,Nayak}.
Thus it is necessary to find a new critical field theory based on
the confinement scenario. In addition, in the critical
theory\cite{Senthil} the conduction electrons are considered to be
bystanders but the spinons are not. Only the spinons and holons
participate in the critical field theory. It seems that this is
not appropriate to both conduction electrons and spinons. When we
approach the quantum critical point from the strong Kondo coupling
regime, critical fluctuations of the Kondo singlets are expected
to affect both the conduction electrons and the spinons. Thus it
seems to be natural that the critical theory should include not
only the spinons but also the conduction electrons. Considering
the above discussion, we can find natural conditions which the
critical theory in the two dimensional Kondo problem should
satisfy. The critical field theory should be written in terms of
internal charge neutral particles as a result of the confinement.
It has to include both the conduction electrons and spinons.
Admitting the two conditions, we can construct the internal charge
neutral particles in terms of the conduction electrons and
spinons. Moreover, in order to explain the non-Fermi liquid
behavior it is natural to consider long range gauge interactions
between the internal charge neutral
particles\cite{Tsvelik,Nagaosa,Ioffe,C_v1,C_v2,Gauge_NFL}. We note
that the gauge interaction here has nothing to do with the
internal gauge field in the U(1) slave boson representation.

Our strategy is firstly to integrate over the holon field
representing fluctuations of the Kondo singlets and to obtain an
effective Lagrangian in terms of the conduction electrons and
spinons. We approach the quantum critical point from the strong
Kondo coupling phase, that is, the Kondo singlet paramagnetism. We
will consider fluctuations of the phase of the holon field only,
i.e., $b_{r} = b_{0}e^{i\phi_{r}}$ with an amplitude $b_0 =
\frac{J_K}{2}|<c_{r\sigma}^{\dagger}f_{r\sigma}>|$. The holon
field carries both an internal charge and an electric charge
associated with the internal U(1) gauge field $a_{\mu}$ and the
electromagnetic field $A_{\mu}$, respectively. The idea is to
divide the phase field into partitions carrying the internal
charge and the electric charge respectively. We rewrite the phase
field as $\phi_{r} = \phi_{fr} + \phi_{cr}$. Here $\phi_{fr}$
carries only internal gauge charge and $\phi_{cr}$, only electric
charge. In order to treat the coupling
$b_{0}e^{-i\phi_{r}}f_{r\sigma}c_{r\sigma}^{\dagger}$ in Eq. (2)
we consider the following gauge transformation \bqa &&
\tilde{f}_{r\sigma} = e^{-i\phi_{fr}}f_{r\sigma} , \nn &&
\tilde{c}_{r\sigma} = e^{i\phi_{cr}}c_{r\sigma} . \eqa After
performing the gauge transformation, we obtain
$b_{0}\tilde{f}_{r\sigma}\tilde{c}_{r\sigma}^{\dagger}$. Phase
fluctuations of the holon field no longer couples to the
conduction electrons and spinons in this term. Instead the
coupling appears in the kinetic term [Eq. (5)]. It is easy to
treat this coupling as we discuss below. The phase field
$\phi_{f(c)}(r)$ satisfies $\nabla\times\nabla\phi_{f(c)}(r) =
2\pi{q_R}\delta(r-R)$ where $q_R$ is a vortex charge (integer) and
$R$, the vortex center. This implies that in the case of non-zero
$q_R$ there exist points where $\phi_{f(c)}(r)$ is not well
defined, i.e., singular. This transformation is usually called a
singular gauge transformation\cite{Tesanovic}. The phase fields
$\phi_{fr}$ and $\phi_{cr}$ are not pure gauge degrees of freedom,
and thus vortex configurations ($q_{R} \not= 0$) should be allowed
physically. As will be discussed below, the critical field theory
is obtained by treating the vortex fields instead of the holon
fields. In the present paper both the phase fields $\phi_{fr}$ and
$\phi_{cr}$ will be taken into account. It should be noted that
there is another singular gauge transformation of $\phi_{fr} =
\phi_{r}$ and $\phi_{cr} = 0$. Utilizing this gauge
transformation, we can also obtain a critical theory near the
quantum critical point. In appendix A we show that this critical
theory Eq. (A5) is totally different from our critical theory Eq.
(11). We find that the critical theory Eq. (A5) cannot explain the
non-Fermi liquid behavior in the Kondo system.

In the singular gauge transformation Eq. (4) we attach
$e^{-i\phi_{fr}}$ to the spinon. As a result the internal gauge
charge of the spinon is screened or neutralized by the phase. The
renormalized spinon $\tilde{f}_{r\sigma}$ carries no internal
gauge charge. This result can be interpreted as follows. As
discussed earlier, strong gauge fluctuations $a_{\mu}$ are
expected to cause the confinement of the internal gauge charges in
$(2+1)D$\cite{Fradkin,Kim_PRL}. The gauge fluctuations confine the
spinons and holons, resulting in the renormalized spinons which
are neutral for the internal gauge charge. This point of view is
analogous to that of Ref. \cite{Kim_PRL,Kim} in the context of
high $T_c$ cuprates. As will be seen below, the internal U(1)
gauge field $a_{\mu}$ can be integrated out exactly in the low
energy limit without affecting the neutralized spinons. The
renormalized conduction electron $\tilde{c}_{r\sigma}$ is also
neutralized. As we approach the quantum critical point,
fluctuations of the hybridization between the conduction electrons
and spinons are very strong. These fluctuations are expected to
result in the renormalization of the conduction electrons.
Although both the renormalized electrons and spinons are
electrically neutral, it will be shown that both particles are
coupled to the electromagnetic field $A_{\mu}$ in our critical
field theory [Eq. (11)].

We rewrite the above Lagrangian [Eq. (2)] in terms of the
renormalized fields [Eq. (4)]   \bqa && L =
\sum_{r}\tilde{c}_{r\sigma}^{\dagger}(\partial_{\tau} -
i\partial_{\tau}\phi_{cr} -
\epsilon({\sqrt{-(\nabla-i\nabla\phi_{cr})^2}}))\tilde{c}_{r\sigma}
\nn && + \sum_{r}\tilde{f}_{r\sigma}^{\dagger}(\partial_{\tau} +
i\partial_{\tau}\phi_{fr} - ia_{\tau})\tilde{f}_{r\sigma} \nn && -
\chi_{0}\sum_{<r,r'>}(\tilde{f}^{\dagger}_{r\sigma}e^{-i(\nabla\phi_{fr}-a_{rr'})}\tilde{f}_{r'\sigma}
+ h.c.) \nn && -
\sum_{r}(b_{0}\tilde{f}_{r\sigma}\tilde{c}_{r\sigma}^{\dagger} +
h.c.) + \sum_{r}\frac{4b_0^2}{J_K}  . \eqa Here we have rewritten
the kinetic energy of the conduction electrons in real space. As a
result of the singular gauge transformation [Eq. (4)] phase
fluctuations of the holon fields are now coupled to both the
renormalized electrons and spinons in the kinetic energy terms. As
will be seen below, quantum fluctuations of the Kondo singlets
result in long range gauge interactions between the two particles
near the quantum critical point. In other words, the phase fields
$\phi_{cr}$ and $\phi_{fr}$ are reinterpreted as two kinds of U(1)
gauge fields.

We introduce two kinds of U(1) gauge fields usually called the
"Doppler" gauge field $v_{\mu}$ and the "Berry" gauge field
$a_{\mu}^{B}$ in the context of the quantum disordered $d-wave$
superconductor for high $T_c$ cuprates\cite{Tesanovic} \bqa &&
v_{\mu} = \frac{1}{2}(\partial_{\mu}\phi_{fr} +
\partial_{\mu}\phi_{cr}) , \nn && a_{\mu}^{B} = \frac{1}{2}(\partial_{\mu}\phi_{fr}
- \partial_{\mu}\phi_{cr}) . \eqa In the context of the quantum
disordered $d-wave$ superconductor the Doppler gauge field
represents supercurrent fluctuations and causes the Doppler shift
in a quasiparticle spectrum. The Berry gauge field represents
vortex fluctuations and encodes the Aharnov-Bohm phase when the
quasiparticles are turning around the vortices. These two gauge
fields are introduced in order to treat strong phase fluctuations
of the Cooper pair. Our Doppler and Berry gauge fields are not
related with those in the superconductor. Instead fluctuations of
the holon field can be easily treated by introducing the two gauge
fields, as will be seen below.

Representing the above Lagrangian [Eq. (5)] in terms of the two
gauge fields [Eq. (6)], we obtain the following Lagrangian in the
continuum limit \bqa && Z
=\int{D\tilde{c}_{\sigma}}{D\tilde{f}_{\sigma}}{Dv_{\mu}}{Da_{\mu}^{B}}{Da_{\mu}}
e^{-\int{d^3x} {\cal L}} , \nn && {\cal L} =
\tilde{c}_{\sigma}(\partial_{\tau}  - iv_{\tau} + ia_{\tau}^{B} +
iA_{\tau} )\tilde{c}_{\sigma} \nn && +
\frac{1}{2m_{c}}|(\nabla-i{\bf v}+i{\bf a}^{B} + i{\bf A}
)\tilde{c}_{\sigma}|^2 \nn && + \tilde{f}_{\sigma}(\partial_{\tau}
+ iv_{\tau} + ia_{\tau}^{B} - ia_{\tau})\tilde{f}_{\sigma} \nn &&
+ \frac{1}{2m_{f}}|(\nabla + i{\bf v} + i{\bf a}^{B} - i{\bf a}
)\tilde{f}_{\sigma}|^2 \nn && -
(b_{0}\tilde{f}_{r\sigma}\tilde{c}_{r\sigma}^{\dagger} + h.c.) +
\frac{4b_0^2}{J_K} \nn && + \frac{\rho}{2}|\partial_{\mu}\phi -
a_{\mu} - A_{\mu}|^2 - i\lambda_{c}(\partial\times{v} -
\partial\times{a}^{B} - J_{cV}) \nn && -
i\lambda_{f}(\partial\times{v} + \partial\times{a}^{B} - J_{fV})
\eqa with $m_c$, the mass of the conduction electron and $m_f \sim
\chi_{0}^{-1}$, that of the spinon. Here the kinetic energy of the
renormalized conduction electrons is explicitly written in the
usual non-relativistic form with an electromagnetic field $A_{\mu}
= (A_{\tau}, {\bf A})$. The low energy Lagrangian of the holon
field is also explicitly shown with the constraints given by two
Lagrange multipliers $\lambda_{c}$ and $\lambda_{f}$. Here $J_{cV}
= \partial\times\partial\phi_{c}$ and $J_{fV} =
\partial\times\partial\phi_{f}$ are the vortex currents of the
holon partitions carrying the electric charge and the internal
charge, respectively. These constraints are used in order to
impose Eq. (6). Remember that the phase of the holon field is
rewritten in the partitions, i.e., $\phi = \phi_c + \phi_f$.
$\rho$ is a stiffness parameter proportional to the condensation
amplitude $b_{0}^{2}$.

Now we integrate over phase fluctuations of the holon field at the
quantum critical point. Performing a standard duality
transformation of the holon Lagrangian in Eq. (7), we obtain an
effective vortex Lagrangian for each partition ($\phi_c$ and
$\phi_f$) of the holon field \bqa && {\cal L}^{dual}_{\phi} =
|(\partial_{\mu} - i\tilde{\lambda}_{c\mu})\Phi_{c}|^2 +
m^{2}|\Phi_{c}|^2 + \frac{u}{2}|\Phi_{c}|^{4} \nn && +
|(\partial_{\mu} - i\tilde{\lambda}_{f\mu})\Phi_{f}|^2 +
m^{2}|\Phi_{f}|^2 + \frac{u}{2}|\Phi_{f}|^{4}  \nn && -
i(\tilde{\lambda}_{c} +
\tilde{\lambda}_{f})_{\mu}(\partial\times{v})_{\mu} +
i(\tilde{\lambda}_{c} -
\tilde{\lambda}_{f})_{\mu}(\partial\times{a}^{B})_{\mu} \nn && +
\frac{1}{2\rho}|\partial\times{c}|^2 +
i(\partial\times{c})_{\mu}(a_{\mu} + A_{\mu} - 2v_{\mu}) . \eqa
Here $\Phi_{c}$ and $\Phi_{f}$ are the vortex fields of $\phi_c$
and $\phi_f$ respectively, and $c_{\mu}$, the vortex gauge field.
$\tilde{\lambda}_{c\mu}$ and $\tilde{\lambda}_{f\mu}$ result from
the shifted Lagrange multipliers $\tilde{\lambda}_{c\mu} =
\lambda_{c\mu} + c_{\mu}$ and $\tilde{\lambda}_{f\mu} =
\lambda_{f\mu} + c_{\mu}$, respectively. $m$ is a vortex mass and
$u$, a coupling strength of self-interaction. The vortex mass is
given by $m^2 \sim J_{K} - J_{K}^{c}$ where $J_{K}^{c}$ is the
critical Kondo coupling. When $J_{K} > J_{K}^{c}$, the vortex
vacuum is energetically favorable. This corresponds to the strong
Kondo coupling phase where the holons are condensed. In the
opposite case vortex condensation is expected to occur. This
corresponds to the weak Kondo coupling phase where the holons are
not condensed. The quantum critical point is obtained at $J_{K} =
J_{K}^{c}$. At the quantum critical point the hybridization
between the conduction electrons and spinons is strongly
fluctuating. This is represented by critical phase fluctuations of
the holon field. In the vortex Lagrangian Eq. (8) $J_{K} =
J_{K}^{c}$ leads the vortex mass to be zero. As a result critical
phase fluctuations of the holon field are represented by critical
vortex fluctuations. The original vortex gauge field $c_{\mu}$ is
decoupled from the vortex fields. Instead the two fields
$\tilde{\lambda}_{c\mu}$ and $\tilde{\lambda}_{f\mu}$ act as the
vortex gauge fields mediating interactions between the vortices. A
similar and detailed derivation in the context of high $T_c$
cuprates can be found in Ref. \cite{Kim}. After integrating over
the vortex gauge field $c_{\mu}$, we obtain a mass term
$\frac{\rho}{2}|a_{\mu} + A_{\mu} - 2v_{\mu} |^2$. The mass is a
relevant parameter in the renormalization group sense. Thus, in
the low energy limit we can set $v_{\mu} = \frac{1}{2}(a_{\mu} +
A_{\mu})$.

Inserting the Doppler gauge field $v_{\mu} = \frac{1}{2}(a_{\mu} +
A_{\mu})$ into Eq. (7) and Eq. (8), and integrating over the
critical vortex fluctuations and the effective vortex gauge fields
$\tilde{\lambda}_{c\mu}, \tilde{\lambda}_{f\mu}$ in Eq. (8) at the
quantum critical point $J_{K} = J_{K}^{c}$, we obtain an effective
Lagrangian \bqa && {\cal L} = \tilde{c}_{\sigma}(\partial_{\tau} +
ia_{\tau}^{B} - \frac{i}{2}a_{\tau} + \frac{i}{2}A_{\tau}
)\tilde{c}_{\sigma} \nn && + \frac{1}{2m_{c}}|(\nabla + i{\bf
a}^{B} - \frac{i}{2}{\bf a} + \frac{i}{2}{\bf A}
)\tilde{c}_{\sigma}|^2 \nn && + \tilde{f}_{\sigma}(\partial_{\tau}
+ ia_{\tau}^{B} - \frac{i}{2}a_{\tau} +
\frac{i}{2}A_{\tau})\tilde{f}_{\sigma} \nn && +
\frac{1}{2m_{f}}|(\nabla + i{\bf a}^{B} - \frac{i}{2}{\bf a} +
\frac{i}{2}{\bf A} )\tilde{f}_{\sigma}|^2 \nn && +
\frac{1}{2g_{B}^2}(\partial\times{a}^{B})\frac{1}{\sqrt{-\partial^2}}(\partial\times{a}^{B})
\nn && + \frac{1}{2g_{A}^2}(\partial\times{a} + \partial\times{A}
)\frac{1}{\sqrt{-\partial^2}}(\partial\times{a} +
\partial\times{A}) \nn && -
(b_{0}\tilde{f}_{r\sigma}\tilde{c}_{r\sigma}^{\dagger} + h.c.) +
\frac{4b_0^2}{J_K}  .  \eqa Here the effective couplings
$g_{B}^{2} = \Bigl(\frac{8}{N_c} + \frac{8}{N_f} \Bigr)^{-1}$ and
$g_{A}^{2} = \Bigl(\frac{2}{N_c} + \frac{2}{N_f} \Bigr)^{-1}$ are
obtained in the usual $1/N$ expansion\cite{Kim_PRL,Kim} where
$N_{c(f)}$ is the flavor number of the vortex $\Phi_{c(f)}$. In
the present case we have $N_c = N_f = 1$. Anomalous kinetic terms
for both the Berry gauge field and internal gauge field result
from critical vortex fluctuations. Shifting the internal U(1)
gauge field $a_{\mu}$ to $\tilde{a}_{\mu} = a_{\mu} + A_{\mu}$ and
the Berry gauge field $a_{\mu}^{B}$ to $\tilde{a}_{\mu}^{B} =
a_{\mu}^{B} - \frac{1}{2}\tilde{a}_{\mu}$ in Eq. (9), we obtain
\bqa && {\cal L} = \tilde{c}_{\sigma}(\partial_{\tau} +
i\tilde{a}_{\tau}^{B} + iA_{\tau} )\tilde{c}_{\sigma} +
\frac{1}{2m_{c}}|(\nabla + i\tilde{{\bf a}}^{B} + i{\bf A}
)\tilde{c}_{\sigma}|^2 \nn && + \tilde{f}_{\sigma}(\partial_{\tau}
+ i\tilde{a}_{\tau}^{B} + iA_{\tau})\tilde{f}_{\sigma} +
\frac{1}{2m_{f}}|(\nabla + i\tilde{{\bf a}}^{B} + i{\bf A}
)\tilde{f}_{\sigma}|^2 \nn && +
\frac{1}{2g_{B}^2}(\partial\times{\tilde{a}}^{B} +
\frac{1}{2}\partial\times{\tilde{a}}
)\frac{1}{\sqrt{-\partial^2}}(\partial\times{\tilde{a}}^{B} +
\frac{1}{2}\partial\times{\tilde{a}}) \nn && +
\frac{1}{2g_{A}^2}(\partial\times{\tilde{a}}
)\frac{1}{\sqrt{-\partial^2}}(\partial\times{\tilde{a}}) \nn && -
(b_{0}\tilde{f}_{r\sigma}\tilde{c}_{r\sigma}^{\dagger} + h.c.) +
\frac{4b_0^2}{J_K}  .  \eqa Now the internal U(1) gauge field
$a_{\mu}$ is decoupled from the renormalized spinon field. This
result seems to be quite natural because the renormalized spinon
is neutral under the internal gauge field. We can easily check
whether the local gauge symmetry in the original Lagrangian Eq.
(2) is preserved in all previous steps. The Lagrangian Eq. (2) has
a $U_{a}(1)\times{U}_{A}(1)$ local gauge symmetry in association
with the internal and electromagnetic U(1) gauge fields
respectively. We note that the local gauge symmetry in Eq. (2) is
preserved in all steps.

Integrating over the internal U(1) gauge field $\tilde{a}_{\mu}$
in Eq. (10), we finally obtain a critical field theory in terms of
the renormalized conduction electrons and renormalized spinons
interacting via the effective U(1) gauge field
$\tilde{a}_{\mu}^{B}$ \bqa && {\cal L} =
\tilde{c}_{\sigma}(\partial_{\tau} + i\tilde{a}_{\tau}^{B} +
iA_{\tau} )\tilde{c}_{\sigma} + \frac{1}{2m_{c}}|(\nabla +
i\tilde{{\bf a}}^{B} + i{\bf A} )\tilde{c}_{\sigma}|^2 \nn && +
\tilde{f}_{\sigma}(\partial_{\tau} + i\tilde{a}_{\tau}^{B} +
iA_{\tau})\tilde{f}_{\sigma} + \frac{1}{2m_{f}}|(\nabla +
i\tilde{{\bf a}}^{B} + i{\bf A} )\tilde{f}_{\sigma}|^2 \nn && +
\frac{1}{2\tilde{g}^2}(\partial\times{\tilde{a}}^{B}
)\frac{1}{\sqrt{-\partial^2}}(\partial\times{\tilde{a}}^{B} ) \nn
&& - (b_{0}\tilde{f}_{r\sigma}\tilde{c}_{r\sigma}^{\dagger} +
h.c.) + \frac{4b_0^2}{J_K}    \eqa with an effective coupling
strength $\tilde{g}^{2} = g_{B}^{2} + \frac{1}{4}g_{A}^2$.
Physically it is clear that the massless effective U(1) gauge
field $\tilde{a}_{\mu}^{B}$ originates from critical fluctuations
of the Kondo singlets (critical phase fluctuations of the holons)
at the quantum critical point. Mathematically this new emergent
gauge structure results from the composite structure in the phase
of the holon field, i.e., $\phi_r = \phi_{fr} + \phi_{cr}$. The
phase $\phi_r$ is invariant under the transformation $\phi_{fr}
\rightarrow \phi_{fr} + \vartheta$ and $\phi_{cr} \rightarrow
\phi_{cr} - \vartheta$. It is easy to verify that the critical
field theory Eq. (11) has a new local gauge symmetry. Under the
gauge transformation $\phi_{fr} \rightarrow \phi_{fr} + \vartheta$
and $\phi_{cr} \rightarrow \phi_{cr} - \vartheta$ the renormalized
electrons and spinons are also transformed by $\tilde{c}'_{\sigma}
= e^{-i\vartheta}\tilde{c}_{\sigma}$ and $\tilde{f}'_{\sigma} =
e^{-i\vartheta}\tilde{f}_{\sigma}$, respectively. If the gauge
field $\tilde{a}_{\mu}^{B}$ is transformed by
$\tilde{a}'^{B}_{\mu} = \tilde{a}_{\mu}^{B} +
\partial_{\mu}\vartheta$, this symmetry is preserved. This
new gauge symmetry dictates a new U(1) gauge field
$\tilde{a}_{\mu}^{B}$.

A critical field theory in the two dimensional Kondo lattice model
is required to satisfy three conditions as discussed earlier.
First the theory should consist of internal charge neutral
particles as a result of confinement. Second it should be the case
that not only the spinons but also the conduction electrons
participate in the critical theory. Third the critical theory
should explain the observed non-Fermi liquid behavior near the
quantum critical point\cite{Si,Senthil}. A possible candidate is
to introduce long range gauge
interactions\cite{Tsvelik,Nagaosa,Ioffe,C_v1,C_v2,Gauge_NFL}
between the internal charge neutral particles. The critical field
theory Eq. (11) satisfies the above three conditions. Eq. (11) is
written in terms of both the conduction electrons and spinons.
They interact via long range gauge interactions mediated by the
U(1) gauge field $\tilde{a}_{\mu}^{B}$. Both particles are
renormalized to be neutral under internal charge.

Now we discuss physics of the two dimensional Kondo lattice model
based on the confinement scenario of the present paper. In the
strong Kondo coupling regime of $J_{K} > J_{K}^{c}$ the conduction
electrons and spinons are strongly hybridized. This is represented
by the holon condensation as discussed earlier. In this case
fluctuations of the Kondo singlets are suppressed. In the
effective Lagrangian of the holon vortex field [Eq. (8)] $J_{K} >
J_{K}^{c}$ leads to a positive vortex mass. As a result the
condensation of the holons are represented by the vacuum of the
vortices. In the vortex vacuum all gauge fields, $a_{\mu}$,
$v_{\mu}$, and $a_{\mu}^{B}$, become massive. This is consistent
with the suppression of Kondo singlet fluctuations. Thus usual
mean field treatment is available. Fermi liquid behavior is
expected owing to the suppression of gauge
fluctuations\cite{Senthil}. In the context of the gauge theory
this phase corresponds to the Higgs-confinement
phase\cite{Fradkin}. It is natural that only neutral particles for
the internal gauge charge appear in an effective field theory. The
renormalized spinons and conduction electrons are neutral under
the internal gauge charge.

Approaching the quantum critical point from the paramagnetic
phase, fluctuations of the hybridization become strong. These
fluctuations are represented by phase fluctuations of the holon
fields. In the dual vortex formulation [Eq. (8)] $J_{K} =
J_{K}^{c}$ results in massless vortex excitations. Thus critical
phase fluctuations of the holon field are represented by critical
vortex fluctuations as discussed before. These cause both the
internal U(1) gauge field $a_{\mu}$ and the Berry gauge field
$a_{\mu}^{B}$ to be massless. As we have seen above, the internal
U(1) gauge field can be integrated out exactly in the low energy
limit without affecting the two renormalized particles. Only one
effective U(1) gauge field remains in our critical field theory.
It is well known that when the kinetic energy of the gauge field
$\tilde{a}_{\mu}^{B}$ is Maxwellian, the critical field theory Eq.
(11) shows non-Fermi liquid behavior owing to scattering with the
massless gauge
fluctuations\cite{Senthil,Tsvelik,Nagaosa,Ioffe,C_v1,C_v2,Gauge_NFL}.
Specific heat is shown to be proportional to $TlnT$ in $3D$
instead of the standard $\gamma{T}$
behavior\cite{Senthil,C_v1,C_v2} where $T$ is temperature. In $2D$
it is proportional to $T^{2/3}$\cite{Senthil,Tsvelik}, which is
observed in $YbRh_2Si_2$ in the low temperature regime near the
quantum critical point\cite{Senthil,Specific_heat}. Conductivity
is shown to be proportional to $T^{-4/3}$ in $2D$ and $T^{-5/3}$
in $3D$\cite{Senthil,Nagaosa,Ioffe}. These anomalous behaviors are
not captured by usual Fermi liquid observed in the case when the
gauge fluctuations are suppressed via the Anderson-Higgs mechanism
in the strong Kondo coupling regime. However, in the present case
the kinetic energy of the gauge field is not Maxwellian owing to
the critical vortex fluctuations. As far as we know, this
effective field theory Eq. (11) has not been examined before in
the literature. Owing to the non-Maxwellian contribution of the
gauge field $\tilde{a}_{\mu}^{B}$ temperature dependence in the
specific heat and conductivity is expected to change. Here we
consider the specific heat and the self-energy of the renormalized
particles in two dimensions.

Integrating over the renormalized spinon and electron fields in
Eq. (11), we obtain an effective action in terms of the effective
gauge field $\tilde{a}_{\mu}^{B}$. Integration over the gauge
field in the usual random phase approximation results in the
following expression of free energy \bqa && \frac{F}{V} = -
\int\frac{d^Dq}{(2\pi)^{D}}
\int\frac{d\omega}{2\pi}coth\Bigl[\frac{\omega}{2T}\Bigr]
tan^{-1}\Bigl[\frac{ImD^{T}(q,\omega)}{ReD^{T}(q,\omega)}\Bigr] .
\eqa Here $D^{T}(q,\omega)$ is the transverse retarded propagator
of the gauge field $\tilde{a}_{\mu}^{B}$ which is renormalized by
the polarization of the spinons and electrons. $V$ is volume of
the system. This expression is valid for any theory where we sum a
sequence of bubble-type diagrams\cite{Tsvelik}. The retarded gauge
field propagator is given by \bqa && D^{T}(q,\omega) = [
\sqrt{-\omega^{2} + q^{2}} - i\omega\sigma(q) +
\chi_{d}q^{2}]^{-1} . \eqa Here $\sigma(q)$ is the conductivity
resulting from both the spinons and electrons. It is given by
$\sigma(q) = \sigma_{f}(q) + \sigma_{c}(q) = (k_{0f} +
k_{0c}){q}^{-1}$ in the clean limit\cite{Nagaosa}. $k_{0f(c)}$ is
near the Fermi momentum which is about the inverse of the lattice
spacing\cite{Nagaosa}. The term $i\omega\sigma(q)$ describes
dissipation of the gauge field due to quasiparticle excitations.
$\chi_{d} = \chi_{f} + \chi_{c} = (24\pi)^{-1}({m}_{f}^{-1} +
{m}_{c}^{-1})$ is the diamagnetic susceptibility for a free
fermion in two dimensions\cite{Nagaosa}. Here $m_{f(c)}$ is the
mass of the spinon (electron). The non-Maxwell contribution
$\sqrt{-\omega^2 + q^2}$ appears in the kernel $D^{T}(q,\omega)$.
Ignoring the $\chi_{d}q^2$ contribution in the kernel owing to the
non-Maxwell contribution $\sqrt{-\omega^2 + q^2}$, we obtain the
imaginary and real part of the kernel \bqa ImD^{T}(q,\omega) && =
\frac{k_0\omega{q}}{\omega^2(-q^2+k_0^2) + q^4} , \nn
ReD^{T}(q,\omega) && =
\frac{q^2\sqrt{-\omega^2+q^2}}{\omega^2(-q^2+k_0^2) + q^4} \eqa
with $k_0 \equiv k_{0f} + k_{0c}$. Inserting these expressions
into Eq. (12), we obtain the following expression for the free
energy \bqa && \frac{F}{V} = -
\int\frac{d^Dq}{(2\pi)^{D}}\int{d\omega}coth\Bigl[\frac{\omega}{2T}\Bigr]
tan^{-1}\Bigl[\frac{k_0\omega}{q\sqrt{-\omega^2+q^2}}\Bigr] . \eqa
In appendix B we perform the momentum and frequency integrals
explicitly. As a result we find \bqa && \frac{F}{V} =
\frac{F_0}{V} + \frac{k_0}{16\pi^2}\Bigl(\frac{3}{2} -
{lnk_0}\Bigr)T^2 - \frac{k_0}{16\pi^2}T^{2}lnT , \eqa where
$F_0/V$ is large negative depending on energy cutoff. The specific
heat is then obtained to be in two dimensions \bqa C_{V} && =
-\frac{T}{V}\Bigl(\frac{\partial^2F}{\partial{T}^2}\Bigr)_{V} \nn
&& = \frac{k_0lnk_0}{8\pi^2}T + \frac{k_0}{8\pi^2}TlnT . \eqa In
the above expression the logarithmic correction seems to be
natural. In the free energy expression Eq. (15) one can find
$q{\times}tan^{-1}[{k_0\omega}/{q^2}]$ in two dimensions in the
limit of $\omega << q$ [appendix B]. When dynamics of the gauge
field is described by the usual Maxwell kinetic energy, it is
given by $q{\times}tan^{-1}[{k_0\omega}/{q^3}]$\cite{Nagaosa}. In
three dimensions it is given by
$q^2{\times}tan^{-1}[{k_0\omega}/{q^3}]$\cite{Tsvelik}. In Ref.
\cite{Tsvelik} $q^2{\times}tan^{-1}[{k_0\omega}/{q^3}]$ is
approximated to $\sim \omega/q$ in the frequency range of
$(k_0\omega)^{1/3} < q$. In the region of ${k_0\omega}/{q^2} < 1$
we can also approximate $q{\times}tan^{-1}[{k_0\omega}/{q^2}]$ to
$\sim \omega/q$ [appendix B]. Both the expressions have the same
momentum and frequency dependence. Thus we can see that the
unusual dynamics described by the non-Maxwell kinetic energy
results in "three dimensional effect". As a consequence we obtain
a logarithmic correction in the temperature dependence of the
specific heat.

Next we consider the imaginary part in the self-energy of the
renormalized conduction electron. In the present formulation the
single electron propagator $G_{\sigma\sigma}(r,\tau) =
<T_{\tau}[c_{\sigma}(r,\tau)c^{\dagger}_\sigma(0,0)]>$ can be
represented in terms of the renormalized conduction electrons \bqa
G_{\sigma\sigma}(r,\tau) && =
<T_{\tau}[c_{\sigma}(r,\tau)c^{\dagger}_\sigma(0,0)]> \nn && =
<T_{\tau}[\tilde{c}_{\sigma}(r,\tau)
e^{i\int_{(0,0)}^{(r,\tau)}{dx_{\mu}}\partial_{\mu}\phi_{c}(x_{\mu})}\tilde{c}^{\dagger}_{\sigma}(0,0)]>
\nn&&=<T_{\tau}[\tilde{c}_{\sigma}(r,\tau)e^{-i\int_{(0,0)}^{(r,\tau)}{dx_{\mu}}\tilde{a}^{B}_{\mu}(x)}
\tilde{c}^{\dagger}_{\sigma}(0,0)]> . \eqa Here we will not
calculate this gauge invariant green's function Eq. (18). Instead
we investigate a simpler one $G_{\sigma\sigma}(r,\tau) =
<T_{\tau}[\tilde{c}_{\sigma}(r,\tau)\tilde{c}^{\dagger}_\sigma(0,0)]>$.
Our objective is to see how the non-Maxwell dynamics of the gauge
field can modify the results in the case of the Maxwell dynamics.
More extensive study of Eq. (18) will be performed in the near
future. Following Ref. \cite{Nagaosa}, we write down the
expression for the imaginary part of the self-energy \bqa
\Sigma_{c}{''}(k,\epsilon_k) && =
\int_{0}^{\infty}{d\omega}\int\frac{d^{D}k'}{(2\pi)^{D}}[n(\omega)
+ 1][1 - f(\epsilon_{k'})] \nn && \times
(k+k')_{\alpha}(k+k')_{\beta}(2m_{c})^{-2} \nn && \times
(\delta_{\alpha\beta} - q_{\alpha}q_{\beta}/q^2)ImD^{T}(q,\omega)
\nn && \times \delta(\epsilon_{k} - \epsilon_{k'} - \omega) . \eqa
Here $q_{\alpha} = (k'-k)_{\alpha}$ is the transfer momentum
between the renormalized electrons and the gauge bosons.
$n(\omega)$ and $f(\omega)$ are the boson and fermion occupation
numbers, respectively. We perform the momentum and frequency
integration explicitly in appendix C. As a consequence we find
\bqa \Sigma_{c}{''}(k,\epsilon_k) &&
=\frac{k_cN_c(0)}{2\pi{m_{c}^{2}}}\int_{0}^{\xi_{k}}d\omega\int_{0}^{\infty}dq
\frac{k_0\omega{q}}{\omega^2(-q^2+k_0^2) + q^4} \nn && \approx
\frac{k_cN_c(0)}{8\pi{m}_{c}^2}\xi_{k} \eqa with $\xi_k =
\epsilon_k - \mu_c$. Here $\epsilon_k$ and $\mu_c$ are the bare
dispersion and chemical potential of the electrons respectively.
$k_c$ and $N_{c}(0)$ are the Fermi momentum and density of states
of the electrons respectively. The last expression $\Sigma_{c}''
\sim \xi_{k}$ is obtained in the limit of $\xi_k << k_0$ [see
appendix C]. In the case of the Maxwell dynamics the self-energy
is found to be $\Sigma'' \sim \xi_k^{2/3}$ in two
dimensions\cite{Nagaosa} and $\Sigma'' \sim \xi_k$ in three
dimensions\cite{C_v2}. Thus we find that the non-Maxwell dynamics
of the gauge field also causes the three dimensional effect to the
self-energy like the case of the specific heat. It is well known
that Fermi liquid shows $\Sigma'' \sim \xi_k^2$. Our critical
field theory Eq. (11) describes non-Fermi liquid near the quantum
critical point.

In the weak Kondo coupling regime of $J_{K} < J_{K}^{c}$ the
hybridization is expected to disappear. The condensation amplitude
of the holon field is expected to vanish, i.e., $b_0 = 0$. In this
case it is not clear how to apply our formulation to this regime.
This is because the singular gauge transformation [Eq. (4)] may be
meaningless owing to the vanishing amplitude. If the condensation
amplitude remained finite like the case of the quantum disordered
superconductor\cite{Tesanovic}, our theory might have been
meaningful. Especially, our theory could be applied to the weak
Kondo coupling regime not far from the quantum critical point. In
this case the holon vortices are condensed and the two gauge
fields, $a_{\mu}$ and $a_{\mu}^{B}$ remain massless. One
difference from the critical field theory is that the kinetic
energy of the gauge fields is Maxwellian. An effective coupling
strength between the renormalized particles and the Berry gauge
field $a_{\mu}^{B}$ is given by the condensation amplitude of the
holon vortex fields\cite{Tesanovic,Kim}.

To summarize, quantum fluctuations of the Kondo singlets result in
renormalization of both the conduction electrons and spinons.
Furthermore, these cause long range gauge interactions between the
renormalized particles near the quantum critical point. As a
consequence the critical field theory is found to consist of the
renormalized electrons and spinons interacting via the new
emergent U(1) gauge field. Investigating the specific heat and the
self-energy of the renormalized electrons, we have found the
non-Fermi liquid behavior near the quantum critical point.

K.-S. Kim especially thanks Dr. Yee, Ho-Ung for correcting English
errors.

\appendix
\section{}
In appendix A we discuss a different singular gauge
transformation. We show that this gauge transformation does not
result in the non-Fermi liquid behavior near the quantum critical
point. We rewrite the singular gauge transformation Eq. (4) \bqa
&& \tilde{f}_{r\sigma} = e^{-i\phi_{fr}}f_{r\sigma} , \nn &&
\tilde{c}_{r\sigma} = e^{i\phi_{cr}}c_{r\sigma} . \eqa If we
choose $\phi_{cr} = 0$ and $\phi_{fr} = \phi_{r}$ with the phase
field $\phi_{r}$ of the holon ($b_{r} = b_0e^{i\phi_{r}}$), we
obtain $\tilde{c}_{r\sigma} = c_{r\sigma}$ and
$\tilde{f}_{r\sigma} = e^{-i\phi_r}f_{r\sigma}$. In this gauge
transformation the electron is the same as before. Only the spinon
is renormalized to carry an electric charge but no internal gauge
charge. Based on this gauge transformation we obtain an effective
Lagrangian \bqa && {\cal L} = {c}_{\sigma}(\partial_{\tau}  +
iA_{\tau} ){c}_{\sigma} + \frac{1}{2m_{c}}|(\nabla + i{\bf A}
){c}_{\sigma}|^2 \nn && + \tilde{f}_{\sigma}(\partial_{\tau} +
i\partial_{\tau}\phi - ia_{\tau})\tilde{f}_{\sigma} +
\frac{1}{2m_{f}}|(\nabla + i\nabla\phi - i{\bf a}
)\tilde{f}_{\sigma}|^2 \nn && -
(b_{0}\tilde{f}_{r\sigma}\tilde{c}_{r\sigma}^{\dagger} + h.c.) +
\frac{4b_0^2}{J_K} + \frac{\rho}{2}|\partial_{\mu}\phi - a_{\mu} -
A_{\mu}|^2  . \eqa Redefining the internal gauge field $a_{\mu}$
as $a_{\mu} = \tilde{a}_{\mu} + \partial_{\mu}\phi_{r}$ (called
the unitary gauge), we obtain the massive internal gauge field
$\tilde{a}_{\mu}$ in the Kondo singlet paramagnetic phase where
the phase of the holon field is coherent ($<e^{i\phi_{r}}> \not=
0$). Thus usual mean field treatment is applicable in this regime.
This is the same result to the effective field theory based on the
singular gauge transformation Eq. (4). See the discussion below
Eq. (11). In the present paper the Kondo singlet paramagnetic
phase is not of our interest. Instead we concentrate on the region
near the quantum critical point of the Kondo lattice model. In
this case it is difficult to apply this unitary gauge owing to
strongly fluctuating holon vortices [see the discussion of the
unitary gauge in the paper of H. Kleinert and F. S. Nogueira,
Nucl. phys. B {\bf 651}, 361 (2003)]. In order to treat this
problem we introduce the Doppler and Berry gauge fields as shown
in Eq. (6). In the gauge transformation of $\phi_{cr} = 0$ and
$\phi_{fr} = \phi_{r}$ in Eq. (A1) the Doppler and Berry gauge
fields are the same $v_{\mu} = a_{\mu}^{B} =
\frac{1}{2}\partial_{\mu}\phi_{r}$.

We rewrite Eq. (A2) in terms of these two gauge fields with
constraints \bqa && {\cal L} = {c}_{\sigma}(\partial_{\tau} +
iA_{\tau} ){c}_{\sigma} + \frac{1}{2m_{c}}|(\nabla + i{\bf A}
){c}_{\sigma}|^2 \nn && + \tilde{f}_{\sigma}(\partial_{\tau} +
iv_{\tau} + ia_{\tau}^{B} - ia_{\tau})\tilde{f}_{\sigma} \nn && +
\frac{1}{2m_{f}}|(\nabla + i{\bf v} + i{\bf a}^{B} - i{\bf a}
)\tilde{f}_{\sigma}|^2 \nn && -
(b_{0}\tilde{f}_{r\sigma}\tilde{c}_{r\sigma}^{\dagger} + h.c.) +
\frac{4b_0^2}{J_K} \nn && + \frac{\rho}{2}|\partial_{\mu}\phi -
a_{\mu} - A_{\mu}|^2 - i\lambda_{c}(\partial\times{v} -
\partial\times{a}^{B}) \nn && - i\lambda_{f}(\partial\times{v} + \partial\times{a}^{B} -
J_{V}) , \eqa where $J_{V} = \partial\times\partial\phi_{r}$ is a
holon vortex three current. Integration over the Lagrange
multipliers $\lambda_{f\mu}$ and $\lambda_{c\mu}$ in Eq. (A3)
recovers Eq. (A2). Now we integrate over the phase field
$\phi_{r}$ of the holon near the quantum critical point. The
methodology is the same as that in the text.

Performing the standard duality transformation of the holon
Lagrangian in Eq. (A3), we obtain an effective Lagrangian for the
holon vortex field \bqa && {\cal L}_{\Phi} = |(\partial_{\mu} -
i\tilde{\lambda}_{f\mu})\Phi|^{2} + m^{2}|\Phi|^{2} +
\frac{u}{2}|\Phi|^{4} \nn && +
\frac{1}{2\rho}|\partial\times{c}|^{2} -
i(\partial\times{c})_{\mu}(a_{\mu} + A_{\mu}) \nn && -
i\lambda_{c}(\partial\times{v} -
\partial\times{a}^{B}) -
i(\tilde{\lambda}_{f\mu} - c_{\mu})(\partial\times{v} +
\partial\times{a}^{B}) . \eqa Here $\Phi$ is the holon vortex
field and $c_{\mu}$, the vortex gauge field.
$\tilde{\lambda}_{f\mu}$ results from the shifted Lagrange
multiplier $\tilde{\lambda}_{f\mu} = \lambda_{f\mu} + c_{\mu}$.
Integration over the vortex gauge field $c_{\mu}$ results in a
mass term for the gauge fields, $\frac{\rho}{2}|v_{\mu} +
a_{\mu}^{B} - a_{\mu} - A_{\mu}|^{2}$. Thus $v_{\mu} + a_{\mu}^{B}
= a_{\mu} + A_{\mu}$ is obtained in the low energy limit.
Integration over the Lagrange multiplier $\lambda_{c\mu}$ gives
$v_{\mu} = a_{\mu}^{B}$. This is what we have expected in the
gauge transformation of $\phi_{cr} = 0$ and $\phi_{fr} = \phi_{r}$
in Eq. (A1). As a result we obtain $a_{\mu} = 2a_{\mu}^{B} -
A_{\mu}$.

Inserting $a_{\mu} = 2a_{\mu}^{B} - A_{\mu}$ and $v_{\mu} =
a_{\mu}^{B}$ into Eq. (A3) and Eq. (A4), we obtain an effective
Lagrangian in the two dimensional Kondo lattice model \bqa &&
{\cal L} = {c}_{\sigma}(\partial_{\tau} + iA_{\tau} ){c}_{\sigma}+
\frac{1}{2m_{c}}|(\nabla + i{\bf A} ){c}_{\sigma}|^2 \nn && +
\tilde{f}_{\sigma}(\partial_{\tau} + iA_{\tau})\tilde{f}_{\sigma}
+ \frac{1}{2m_{f}}|(\nabla + i{\bf A} )\tilde{f}_{\sigma}|^2 \nn
&& - (b_{0}\tilde{f}_{r\sigma}\tilde{c}_{r\sigma}^{\dagger} +
h.c.) + \frac{4b_0^2}{J_K} \nn && + |(\partial_{\mu} -
i\tilde{\lambda}_{f\mu})\Phi|^{2} + m^{2}|\Phi|^{2} +
\frac{u}{2}|\Phi|^{4} \nn && -
i2\tilde{\lambda}_{f\mu}(\partial\times{a}^{B})_{\mu} . \eqa Eq.
(11) in the text can be reduced to this effective Lagrangian in
the case of $\tilde{a}_{\mu}^{B} = 0$ obtained by the gauge
transformation in appendix A. As shown in this effective
Lagrangian, Kondo singlet fluctuations represented by holon vortex
fluctuations are decoupled from the quasiparticles $c_{\sigma}$
and $\tilde{f}_{\sigma}$. Thus mean field like behavior for the
quasiparticles is expected to occur even near the quantum critical
point. We know that this cannot explain the quantum critical
behavior in the Kondo lattice model. This gauge transformation
does not give us satisfactory results.

The problem associated with the singular gauge transformation was
already discussed in the context of the quantum disordered
$d-wave$ superconductor\cite{Tesanovic,Ye}. The gauge
transformation in appendix A is usually called the "Anderson
gauge"\cite{Ye} while that in the text, the "FT
gauge"\cite{Tesanovic}. In the Anderson gauge the phase field of
the Cooper pair living on links of the lattices is attached to
only one kind of electrons with spin $\uparrow$ or spin
$\downarrow$. Consider the Cooper pair term
$|\Delta_0|e^{-i\phi_{rr'}}c_{\uparrow{r}}c_{\downarrow{r'}}$,
where $|\Delta_0|$ and $\phi_{rr'}$ are the amplitude and phase of
the Cooper pair respectively. In the long wave length limit of our
interest we can perform the continuum approximation $a = |r-r'|
\rightarrow 0$ with the lattice spacing $a$. This continuum
approximation is well adopted in the low energy
limit\cite{Tesanovic,Ye}. Then we obtain the expression
$|\Delta_0|e^{-i\phi_{r}}c_{\uparrow{r}}c_{\downarrow{r}}$ for the
Cooper pair term. We consider the following singular gauge
transformation \bqa && \tilde{c}_{\uparrow{r}} =
e^{-i\phi_{\uparrow{r}}}c_{\uparrow{r}} , \nn &&
\tilde{c}_{\downarrow{r}} =
e^{-i\phi_{\downarrow{r}}}c_{\downarrow{r}}  . \eqa Here the phase
fields $\phi_{\uparrow{r}}$ and $\phi_{\downarrow{r}}$ satisfy
$\phi_{r} = \phi_{\uparrow{r}}+\phi_{\downarrow{r}}$. Inserting
these expressions into the Cooper pair term, we obtain
$|\Delta_0|\tilde{c}_{\uparrow{r}}\tilde{c}_{\downarrow{r}}$. The
coupling between the phase of the Cooper pairs and the electrons
now appears in the kinetic energy term of the
electrons\cite{Tesanovic,Ye}. In Eq. (A6) $\phi_{\uparrow{r}} =
\phi_{r}$ and $\phi_{\downarrow{r}} = 0$ is the Anderson
gauge\cite{Ye}. On the other hand, in the FT gauge the phases are
randomly chosen\cite{Tesanovic}. Basically the present Kondo
problem seems to be similar to the problem of the quantum
disordered superconductor. In the Anderson gauge dynamics of the
Cooper pairs is also decoupled from that of the BCS
quasiparticles\cite{Ye} like the above effective Lagrangian Eq.
(A5). It is difficult to determine which is better a priori. The
theory that explains experiments well may be considered as the
correct one. But we can argue that both the spinons and conduction
electrons have the same position. Thus Kondo singlet fluctuations
are expected to cause the same effect to both the spinons and
electrons. Our effective field theory Eq. (11) shows this
symmetric property.

\section{}

In appendix B we perform the momentum and frequency integrals in
Eq. (15) to obtain Eq. (16). We concentrate on quantum
fluctuations in Eq. (15), i.e., $\omega
> T$, so that $coth\Bigl[\frac{\omega}{2T}\Bigr]$ may be replaced
by unity\cite{Ubbens}. We obtain the following expression in two
dimensions \bqa \frac{F}{V} &&
=-\int\frac{d^{D}q}{(2\pi)^{D}}\int_{0}^{\infty}\frac{d\omega}{2\pi}coth\Bigl[\frac{\omega}{2T}\Bigr]
tan^{-1}\Bigl[\frac{k_0\omega}{q\sqrt{-\omega^2+q^2}}\Bigr] \nn &&
= -
\frac{1}{4\pi^{2}}\int_{0}^{\infty}{dq}q\int_{0}^{\infty}{d\omega}coth\Bigl[\frac{\omega}{2T}\Bigr]
tan^{-1}\Bigl[\frac{k_0\omega}{q\sqrt{-\omega^2+q^2}}\Bigr] \nn &&
\approx -
\frac{1}{4\pi^{2}}\int_{0}^{\infty}{dq}q\int_{T}^{v_{c}q}{d\omega}
tan^{-1}\Bigl[\frac{k_0\omega}{q\sqrt{-\omega^2+q^2}}\Bigr] . \eqa
Here $v_{c}$ is the Fermi velocity of the conduction electron. We
are interested in how this quantity depends on the lower cutoff
$T$ in the frequency integral. For small frequencies $\omega <<
v_{c}k_{c}$ with the electron Fermi momentum $k_{c}$, the $q$
integration can be done first\cite{Ubbens}. In addition, in this
limit we can approximate
$tan^{-1}\Bigl[{k_0\omega}/{q\sqrt{-\omega^2+q^2}}\Bigr]$ to
$tan^{-1}\Bigl[{k_0\omega}/{q^2}\Bigr]$. Introducing the energy
cutoff $\omega_c < v_{c}k_{c}$, we find the free energy as a
function of temperature \bqa \frac{F}{V} && \approx -
\frac{1}{4\pi^{2}}\int_{T}^{\omega_c}{d\omega}\int_{0}^{\infty}{dq}q
tan^{-1}\Bigl[\frac{k_0\omega}{q^2}\Bigr] \nn && =  -
\frac{k_0}{8\pi^{2}}\int_{T}^{\omega_c}{d\omega}\Bigl[(1-lnk_0)\omega
- \omega{ln}\omega \Bigr] \nn && = -
\frac{k_0}{16\pi^2}\Bigl(\frac{3}{2}-ln(k_0\omega_c)\Bigr)\omega_{c}^{2}
\nn && + \frac{k_0}{16\pi^2}\Bigl(\frac{3}{2} - {lnk_0}\Bigr)T^2 -
\frac{k_0}{16\pi^2}T^{2}lnT . \eqa This momentum and frequency
integration was done by Mathematica 5.0.

The above integration can be performed analytically in an
appropriate approximation. If we concentrate on the frequency
range $k_0\omega/q^2 < 1$, the $tan^{-1}[{k_0\omega}/{q^2}]$
function can be approximated to ${k_0\omega}/{q^2}$. This
approximation is also used in Ref. \cite{Tsvelik}. As a result we
obtain the following expression of the free energy \bqa
\frac{F}{V} && \approx -
\frac{1}{4\pi^{2}}\int_{T}^{\omega_c}{d\omega}\int_{0}^{\infty}{dq}q
tan^{-1}\Bigl[\frac{k_0\omega}{q^2}\Bigr] \nn && \approx -
\frac{1}{4\pi^{2}}\int_{T}^{\omega_c}{d\omega}\int_{\sqrt{k_0\omega}}^{k_0}{dq}
\frac{k_0\omega}{q} \nn && = -
\frac{k_0}{8\pi^{2}}\int_{T}^{\omega_c}{d\omega}\Bigl[\omega{lnk_0}
- \omega{ln}\omega \Bigr] \nn && = -
\frac{k_0}{16\pi^2}\Bigl(\frac{1}{2}+ln(k_0/\omega_c)\Bigr)\omega_{c}^{2}
\nn && + \frac{k_0}{16\pi^2}\Bigl(\frac{1}{2} + {lnk_0}\Bigr)T^2 -
\frac{k_0}{16\pi^2}T^{2}lnT . \eqa This expression is almost the
same as Eq. (B2) except some numerical factors. The specific heat
can be easily evaluated from the above free energy Eq. (B3)  \bqa
C_{V} && = - T\Bigl(\frac{\partial^{2}F}{\partial{T}^2}\Bigr)_{V}
\nn && = \frac{k_0}{8\pi^2}\Bigl(\frac{1}{2}-lnk_0\Bigr)T +
\frac{k_0}{8\pi^2}TlnT . \eqa This expression also has the same
functional form with Eq. (17).

\section{}

In appendix C we calculate the imaginary part in the self-energy
of the conduction electron in the same way of Ref. \cite{Nagaosa}.
We rewrite Eq. (18) \bqa \Sigma_{c}{''}(k,\epsilon_k) && =
\int_{0}^{\infty}{d\omega}\int\frac{d^{D}k'}{(2\pi)^{D}}[n(\omega)
+ 1][1 - f(\epsilon_{k'})] \nn && \times
(k+k')_{\alpha}(k+k')_{\beta}(2m_{c})^{-2} \nn && \times
(\delta_{\alpha\beta} - q_{\alpha}q_{\beta}/q^2)ImD^{T}(q,\omega)
\nn && \times \delta(\epsilon_{k} - \epsilon_{k'} - \omega) . \eqa
As usual, only states $k'$ near the Fermi surface contribute and
it is convenient to introduce the variable $\xi_k = \epsilon_k -
\mu_c$\cite{Nagaosa} \bqa && \Sigma_{c}{''}(k,\epsilon_k) =
\frac{N_c(0)}{2\pi{m}_{c}^2}\int_{0}^{\infty}{d\omega}
\int{d\xi'}{d\theta}\delta(\xi_k-\xi'-\omega) \nn && \times
[n(\omega) + 1][1 - f(\xi')]|{\bf k} \times{\bf\hat{q}}|^2
\frac{k_0\omega{q}}{\omega^2(-q^2+k_0^2) + q^4}  . \eqa Here
$\theta$ is the angle between $k$ and $k'$ and thus the transfer
momentum $q = k' - k$ is given by $q = 2k_csin(\theta/2)$ with the
Fermi momentum $k_c$. $N_c(0) = m_c/2\pi$ is the density of states
of the electrons. In the above the explicit form of the gauge
field propagator Eq. (14) is used. In the present study we
concentrate on the case of $T = 0 K$. Then the above expression
becomes\cite{Nagaosa} \bqa \Sigma_{c}{''}(k,\epsilon_k) &&
=\frac{k_cN_c(0)}{2\pi{m}_{c}^2}\int_{0}^{\xi_k}{d\omega}\int_{0}^{\infty}{dq}
\frac{k_0\omega{q}}{\omega^2(-q^2+k_0^2) + q^4} . \eqa Performing
the momentum and frequency integration analytically, we obtain the
following expression of the self-energy
$\Sigma_{c}{''}(k,\epsilon_k)$ in the limit of $\xi_k << k_0$ \bqa
\Sigma_{c}{''}(k,\epsilon_k) && =
\frac{k_cN_c(0)k_0}{2\pi{m}_{c}^2}\int_{0}^{\xi_k}{d\omega}
\Bigl[\frac{\pi}{2}\frac{1}{\sqrt{4k_0^2-\omega^2}} \nn && +
\frac{1}{\sqrt{4k_0^2-\omega^2}}tan^{-1}\Bigl(\frac{\omega}{\sqrt{4k_0^2-\omega^2}}\Bigr)\Bigr]
\nn && =
\frac{k_cN_c(0)k_0}{4\pi{m}_{c}^2}\Bigl[{\pi}sin^{-1}\Bigl(\frac{\xi_k}{2k_0}\Bigr)
+ \Bigl(sin^{-1}\Bigl(\frac{\xi_k}{2k_0}\Bigr)\Bigr)^{2}\Bigr] \nn
&& \approx \frac{k_cN_c(0)}{8\pi{m}_{c}^2}\xi_{k} . \eqa


\begin{thebibliography}{9}
\bibitem{Gauge_NFL} It is well known that long range gauge interactions can
result in non-Fermi liquid behavior. Quasiparticle pole (the
hallmark of Fermi liquid) in a single particle green's function
disappears owing to scattering with massless gauge
fluctuations\cite{Tsvelik}. Moreover, the gauge interactions cause
anomalous behavior in correlation functions, not captured in the
context of Fermi liquid\cite{Senthil,Nagaosa,Ioffe,C_v1,C_v2}.
\end{thebibliography}
\end{document}